\documentclass[final,3p,times]{elsarticle}

\usepackage{amssymb}
\usepackage{amsmath}

\journal{Journal of Subatomic Particles and Cosmology}
\usepackage{lineno}

\begin{document}

\begin{frontmatter}



\title{Search for $^4\overline{\mathrm{Li}}$ from $\overline{\mathrm{p}}+{}^3\overline{\mathrm{He}}$ femtoscopy in Pb-Pb collisions at $\sqrt{s_\mathrm{NN}} =5.36$~TeV with ALICE}

\author[aaa,bbb]{Giorgio Alberto Lucia on behalf of the ALICE Collaboration}
\affiliation[aaa]{organization={Politecnico di Torino},
             addressline={Corso Duca degli Abruzzi, 24},
             city={Torino},
             postcode={10129},
             country={IT}}

 \affiliation[bbb]{organization={INFN, Sezione di Torino},
             addressline={Via Pietro Giuria, 1},
             city={Torino},
             postcode={10125},
             country={IT}}

\begin{abstract}
In these proceedings, the search for $^4\overline{\mathrm{Li}}$ in Pb-Pb collisions with ALICE is presented. The particle is reconstructed via its two-body decay channel $^4\overline{\mathrm{Li}} \rightarrow \overline{\mathrm{p}} + {}^3\overline{\mathrm{He}}$, using femtoscopic correlations. A dedicated analysis procedure has been developed to subtract the correlated and uncorrelated $\overline{\mathrm{p}} + ^3\overline{\mathrm{He}}$ combinatorial background. The production yield of $^4\overline{\mathrm{Li}}$ is compared with that of $^4\overline{\mathrm{He}}$. Owing to the presence of several $^4\mathrm{Li}$ resonances and their spin degeneracy, the production of $^4\overline{\mathrm{Li}}$ is expected to be enhanced with respect to the compact $^4\overline{\mathrm{He}}$, making this new measurement a sensitive probe for testing nucleosynthesis models in high-energy hadronic collisions.
\end{abstract}




\end{frontmatter}



\section{Introduction}
Shallow-bound (anti)nuclear states are produced under extreme conditions at the LHC. Understanding their production mechanism has been a challenge for the past years: it plays an important role unveiling the internal structure of these objects~\cite{ALICE:2026tqr}, and to interpreting  astrophysical observations. In particular, antinuclei may be produced from dark-matter annihilation, making them potential indirect probes of dark-matter processes~\cite{Ibarra:2012cc}. 

(Anti)$^4\mathrm{Li}$ is a very rare nuclear state consisting of three protons and one neutron. Four resonant states of $^4\mathrm{Li}$ are expected~\cite{Tilley:1992zz}, with similar mass widths, lifetimes of the order of $\tau \sim 10^{-22}$~s, masses differing by a few MeV, and decays proceeding via the two-body channel $\mathrm p+{}^3\mathrm{He}$. Although $^{4}\mathrm{Li}$ was first observed in 1965~\cite{PhysRevLett.15.300}, its internal structure remains unclear, and the antinucleus $^4\overline{\mathrm{Li}}$ has not been observed yet.

In the nucleon-coalescence picture, the nuclear wave function and the spatial extent of $^4\mathrm{Li}$ are key ingredients governing its production rate~\cite{
Bazak:2020wjn}. Recent measurements of other exotic nuclear states, such as the hypernucleus $^3_\Lambda\mathrm H$, demonstrate the sensitivity of production observables to the nuclear structure~\cite{ALICE:2026tqr}.

The ALICE Collaboration has observed several $A=4$ nuclear states~\cite{ALICE:2023qyl, ALICE:2024djx} 
. Their production yields are 
described by thermal models, assuming point-like objects with abundances determined only by their mass and spin. In contrast, the coalescence picture predicts a dependence on the nuclear radius, which is 
a relevant factor for 
spatially extended states. 
Ref.~\cite{Bazak:2020wjn} predicts a strong variation of the $^4\mathrm{Li}/^4\mathrm{He}$ yield ratio with the assumed $^4\mathrm{Li}$ radius, although the prediction is limited by assumptions on the $^4\mathrm{Li}$ wave function. As the first resonant nucleus investigated in heavy-ion collisions, $^4\mathrm{Li}$ provides a unique opportunity to test these production models in the resonant nuclear sector.

In this paper, we present a search for $^4\overline{\mathrm{Li}}$ production in Pb--Pb collisions at $\sqrt{s_\mathrm{NN}} = 5.36$~TeV, based on a sample of $N_\mathrm{events}=1.21\cdot10^9$ events collected by ALICE during LHC Run 3 in the 0-10\% centrality interval. No significant $^4\overline{\mathrm{Li}}$ signal is observed, and an upper limit on its production yield is derived. The result is compared with thermal-model predictions and with the 
measurement of (anti)alpha production in central Pb-Pb collisions at $\sqrt{s_\mathrm{NN}} = 5.02$~TeV~\cite{ALICE:2023qyl}.

\section{Analysis details}
\begin{figure}[h!]
    \centering
    \includegraphics[width=0.54\linewidth]{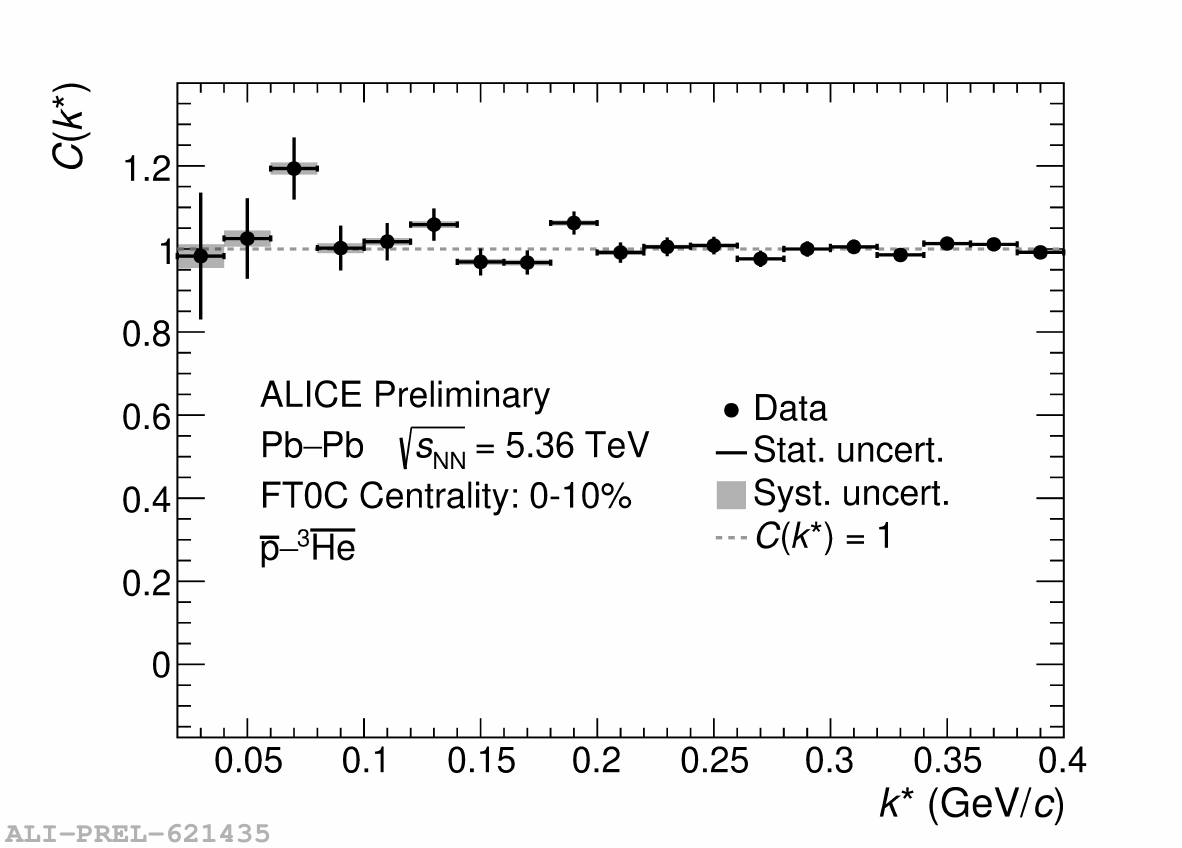}
    \hfill
    \includegraphics[width=0.36\linewidth]{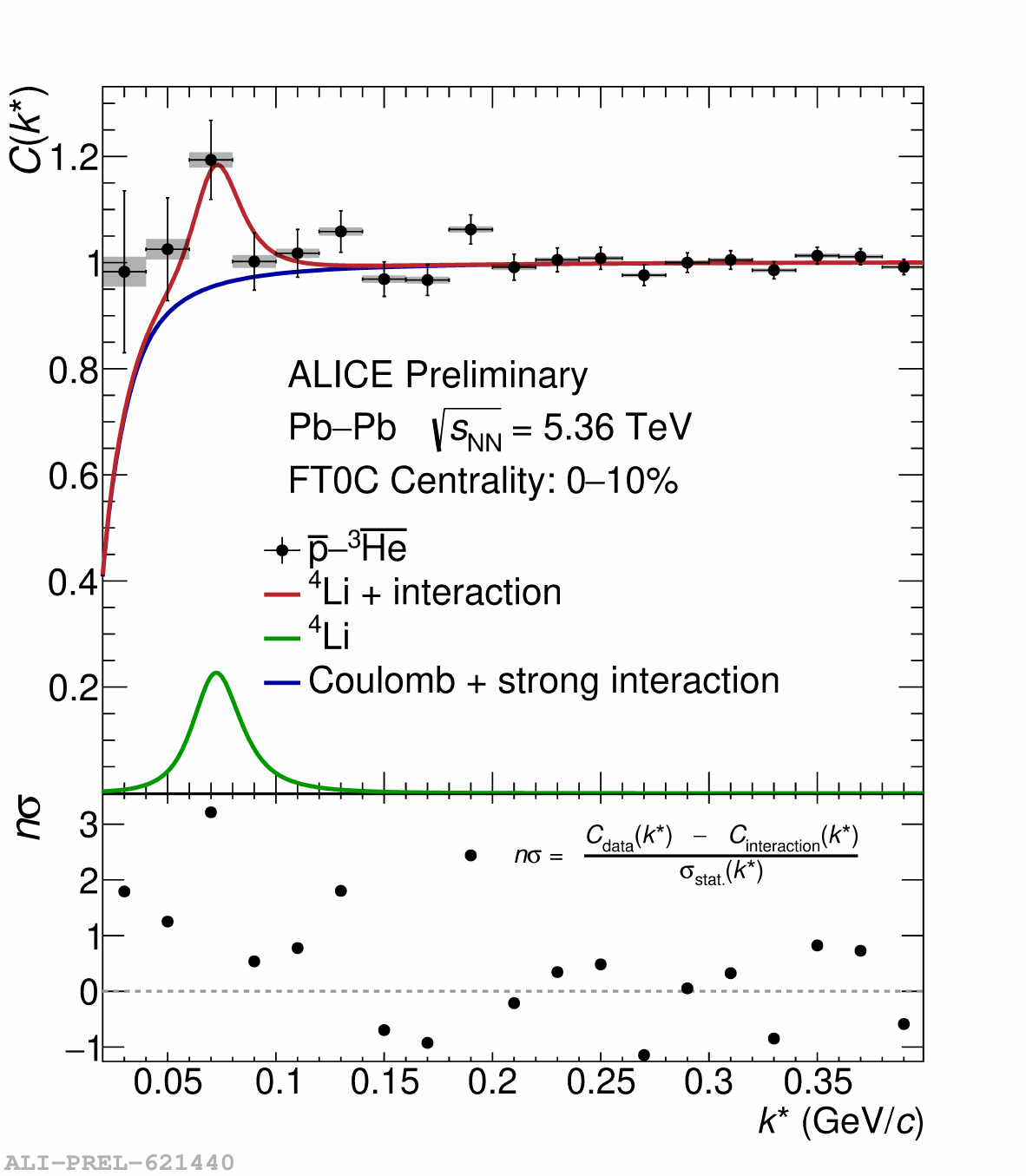}
    \caption{Left panel: femtoscopic correlation function for the $\overline{\mathrm{p}} - ^3\overline{\mathrm{He}}$ pairs in the 0-10\% centrality interval. Black bars and grey boxes represent statistical and systematic uncertainties, respectively. Right panel: Fit to the $\overline{\mathrm{p}}-{}^3\overline{\mathrm{He}}$ correlation function. The interaction contribution, the $^4\overline{\mathrm{Li}}$ resonance contribution, and the total fit are represented by the blue, green, and red curves, respectively. The lower panel shows the 
    difference between the data 
    and the interaction-only contribution, normalised by the statistical uncertainties.}
    \label{fig:data}
\end{figure}

This analysis focuses on the  $^4\overline{\mathrm{Li}} \rightarrow \overline{\mathrm p} + {}^3\overline{\mathrm{He}}$ decay in Pb--Pb collisions at 0-10\% centralities, where the yield is expected to be maximal. The $^4\overline{\mathrm{Li}}$ signal is extracted measuring the femtoscopic correlation function of $\overline{\mathrm p}-{}^3\overline{\mathrm{He}}$ pairs. Femtoscopy techniques probe two-particle interactions by measuring their correlation as a function of the relative momentum in the pair rest frame, $\vec{k}^{\,*}
$~\cite{Lisa:2005dd}. The correlation function is defined by the Koonin--Pratt formula
, the first equality in Eq.~\ref{eq:cf},
\begin{equation}
\label{eq:cf}
    C(\vec{k}^{\,*})=\int S(\vec{r}^{\,*})
    |\psi(\vec{r}^{\,*},\vec{k}^{\,*})|^2\,\mathrm d\vec{r}^{\,*}=\mathcal N\frac{N_\mathrm{SE}(k^*)}{N_\mathrm{ME}(k^*)}\,,
\end{equation}
where $S(\vec{r}^{\,*})$ is the relative spatial distribution of the particle-emitting source, and $\psi(\vec{r}^{\,*},\vec{k}^{\,*})$ is the two-particle wave function, encoding the final-state interactions (FSI). Experimentally, the correlation function is calculated as the second equality in Eq.~\ref{eq:cf}, where $N_\mathrm{SE}$ and $N_\mathrm{ME}$ are the same-event and mixed-event pair distributions, respectively~\cite{Lisa:2005dd}. Mixed-event pairs are formed combining particles from different events with primary vertex positions differing by less than 2~cm and in the same 2.5\%-wide centrality class. The measured correlation function for the $\overline{\mathrm{p}} - {}^3\overline{\mathrm{He}}$ pairs in the 0-10\% centrality interval is shown in the left panel of Fig.~\ref{fig:data}. The normalization factor $\mathcal{N}$ is determined by requiring the correlation function to approach unity in the region $k^*\in[0.2,0.4]$~GeV/$c$, where femtoscopic correlations are expected to be negligible.

To fit the data, a model of the correlation function is constructed (Eq.~\ref{eq:cf-used}), that incorporates the emitting source distribution, the two-particle interactions, and the resonance contribution. 
\begin{equation}
\label{eq:cf-used}
    C(k^*) = A_\mathrm{interaction} [\lambda_{\overline{\mathrm p} + {}^3\overline{\mathrm{He}}} C_{\overline{\mathrm p} + {}^3\overline{\mathrm{He}}}(k^*) + \lambda_{\Sigma^+ + {}^3\overline{\mathrm{He}}} C^\mathrm{Coulomb}_{\overline{\Sigma^+} + {}^3\overline{\mathrm{He}}}(k^*) + (1 - \lambda_{\overline{\mathrm p} + {}^3\mathrm{He}} - \lambda_{\overline{\Sigma^+} + {}^3\overline{\mathrm{He}}})C_\mathrm{uncorr.}(k^*)] + A_{^4\overline{\mathrm{Li}}} C_{^4\overline{\mathrm{Li}}}(k^*)
\end{equation}
The interaction and resonance terms are weighted by the normalisations $A_\mathrm{interaction}$ and $A_{^4\overline{\mathrm{Li}}}$. The source is described by an isotropic Gaussian distribution, whose radius $R_\mathrm s = (7.12 ^{+0.04}_{-0.06})$~fm is determined with the parametrization of Ref.~\cite{ALICE:2025wuy}, using the measured charged-particle multiplicity in Pb--Pb collisions in the 0-10\% centrality interval~\cite{ALICE:2025cjn}. The $\overline{\mathrm{p}}-{}^3\overline{\mathrm{He}}$ FSI term is modelled using CATS~\cite{Mihaylov:2018rva}. CATS natively accounts for the Coulomb interaction, while the strong interaction is described by a square-well potential 
tuned to reproduce the measured $\mathrm p-^3\mathrm{He}$ scattering phase shifts~\cite{Daniels:2010af}. The correlation is also sensitive to pairs initially composed of $^3\overline{\mathrm{He}}$ and $\overline{\Lambda}$ or $\overline{\Sigma^+}$, which in total constitute about 2\% of the sample. After the FSI, the strange baryon decays into a $\overline{\mathrm{p}}$, giving rise to a residual contribution to the measured $\overline{\mathrm p}-{}^3\overline{\mathrm{He}}$ correlation. Since no experimental measurements of the strong interaction between hyperons and $^3\mathrm{He}$ are currently available, only the contribution from the $\overline{\Sigma^+}-{}^3\overline{\mathrm{He}}$ electromagnetic interaction is considered with the term $C^\mathrm{Coulomb}_{\overline{\Sigma^+} + {}^3\overline{\mathrm{He}}}$. The potential impact of the neglected strong interaction is expected to be limited by its short-range nature: for the relatively large source size considered in this analysis, only a small fraction of pairs is expected to reach the region where the strong interaction is relevant. The sample also contains other pairs, such as those including misidentified particles, which are assumed to be uncorrelated, i.e. $C_\mathrm{uncorr.}(k^*) = 1$. In the model, the relative contribution of each pair category is weighted by the $\lambda$ parameters, which account for the particle purities and the fractions of primary and secondary particles, following the procedure used in previous femtoscopy measurements~\cite{ALICE:2025wuy}.
Finally, 
$C_{^4\overline{\mathrm{Li}}}(k^*)$ is obtained by modelling the $k^*$ distribution of pairs originating from the resonance decay. The invariant-mass distribution is described by a Sill function~\cite{Giacosa:2021mbz}, which is transformed into relative-momentum space and divided by the mixed-event distribution, following the procedure described in Eq.~\ref{eq:cf}. The resonance shape is modelled as the ground state, with a mass of $m=3.751$~GeV/$c^2$ and a width of $\Gamma=0.006$~GeV/$c^2$~\cite{Tilley:1992zz}. These values are used as the mass and width parameters of the Sill function. This functional form is used to describe the total resonance contribution, including the excited states, as their expected shapes are similar to that of the ground state.

The fit result is shown in the right panel of Fig.~\ref{fig:data}. The contributions of the interaction, of the $^4\overline{\mathrm{Li}}$ 
, and the total fit are represented by the blue, green, and red curves, respectively. In the lower panel, the pull of the data with respect to the interaction-only contribution highlights the deviation at the $k^*$ values corresponding to the resonance peak. 

\section{Results and discussion}

The statistical significance of the $^4\overline{\mathrm{Li}}$ signal is evaluated by comparing the $\chi^2$ value observed in the data with the distribution obtained from pseudo-experiments generated under the interaction-only hypothesis. Within a $3\sigma$ window around the expected position of the resonance ground state, the resulting significance is $Z = 2.1\sigma$. As the significance of the excess is less than $3\sigma$, a 95\% confidence level upper limit on the production yield is set.

The raw yield $\mu$, used to determine the upper limit, is extracted from the fit result shown in the right panel of Fig.~\ref{fig:data} inverting Eq.~\ref{eq:cf}, applying the normalization used to construct the correlation function. Its statistical uncertainty is evaluated with pseudo-experiments, where the same-event and mixed-event distributions are Poisson-sampled and the full analysis chain is repeated. The resulting yield distribution is fitted with a Gaussian, with parameter $\sigma$, and the upper limit is calculated as $\mu+1.96\sigma$. 

Systematic uncertainties are included by varying the correlation function and its model in the pseudo-experiments. The source radius is varied within its $1\sigma$ uncertainty, the $\lambda$ parameters by 10\%, and the signal selection by repeating the analysis with varied selection criteria. The least stringent resulting upper limit is quoted as a conservative estimate.

\begin{figure}
    \centering
    \includegraphics[width=0.55\linewidth]{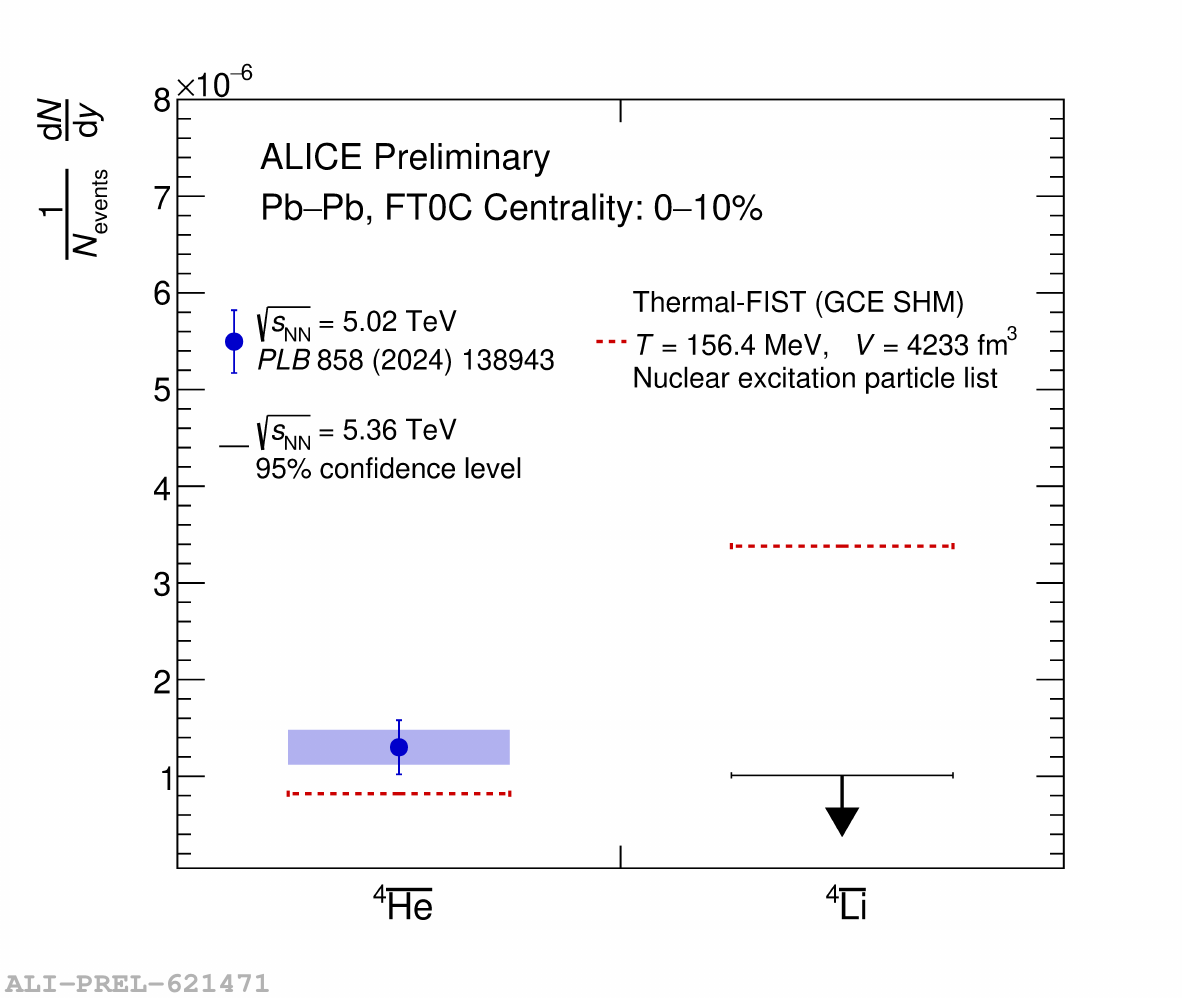}
    \caption{95\% confidence level upper limit to the $^4\overline{\mathrm{Li}}$ production in central Pb-Pb collisions (black arrow), compared to the measured yield of $^4\overline{\mathrm{He}}$ in central Pb-Pb collisions at $\sqrt{s_\mathrm{NN}} = 5.02$~TeV~\cite{ALICE:2023qyl}. The measurements are compared to with Thermal-FIST predictions~\cite{Vovchenko:2019pjl}.}
    \label{fig:plot}
\end{figure}

The upper limit is corrected for the number of analysed events, detector acceptance, reconstruction efficiency, and rapidity acceptance, following the procedure used in previous light-nuclei measurements~\cite{ALICE:2023qyl}. The resulting upper limit in the 0-10\% most central Pb-Pb collisions is
\begin{equation}
    1 / N_\mathrm{events}\,\mathrm dN / \mathrm dy < 1.02 \cdot 10^{-6},
\end{equation}
and is shown in Fig.~\ref{fig:plot}. It is compared with the measured $^4\overline{\mathrm{He}}$ yield in a similar collision system~\cite{ALICE:2023qyl}.

Both results are compared with Thermal-FIST predictions~\cite{Vovchenko:2019pjl}, obtained within the Grand Canonical Ensemble using the temperature and volume parametrization of Ref.~\cite{Vovchenko:2019kes} and the charged-particle multiplicity measured in the 0-10\% centrality interval~\cite{ALICE:2025cjn}. While the predicted $^4\overline{\mathrm{He}}$ yield agrees with the measurement, the prediction for $^4\overline{\mathrm{Li}}$ exceeds the upper limit, suggesting that the thermal model overestimates the production of this state.

On the other hand, two-body coalescence models remain compatible with the upper limit to the $^4\overline{\mathrm{Li}}/^4\overline{\mathrm{He}}$ ratio, although they present a strong dependency on the assumed nuclear wave function and radius for $^4\overline{\mathrm{Li}}$.




\newpage
\bibliographystyle{elsarticle-num}
\bibliography{bib}



\end{document}